\documentclass[10pt]{article}

\usepackage[T1]{fontenc}
\usepackage{lmodern}
\usepackage{microtype}

\usepackage[LGRgreek,frenchmath]{mathastext}
\usepackage[mathscr]{eucal}
\usepackage[sans]{dsfont}
\usepackage{bbold}
\usepackage{bbm}
\usepackage{graphicx}
\usepackage{mdwlist}
\usepackage{natbib}
\setcitestyle{authoryear}
\usepackage[dvipsnames]{xcolor}
\usepackage[plainpages=false, pdfpagelabels, backref=page]{hyperref}
	\hypersetup{
		colorlinks   = true,
		citecolor    = RoyalBlue,
		linkcolor    = RubineRed,
		urlcolor     = MidnightBlue
	}
\usepackage[paperwidth=8.5in,paperheight=11.00in,top=1.00in, bottom=1.00in,
            left=0.75in, right=0.75in]{geometry}
\usepackage{mathtools}
\mathtoolsset{showonlyrefs=true}
\usepackage{amsmath}
\usepackage{amssymb}
\usepackage{amsfonts}
\usepackage{amsthm}
\allowdisplaybreaks
\newtheoremstyle{plain}
  {}{}{\itshape}{}{\mdseries\scshape}{.}{ }
  {\thmname{#1}\thmnumber{ #2}\ifx#3\empty\else\ (#3)\fi}
\theoremstyle{plain}
\newtheorem{theorem}{\underline{Theorem}}

\newtheorem{proposition}[theorem]{\underline{Proposition}}

\newtheoremstyle{definition}
  {}{}{}{}{\mdseries\scshape}{.}{ }
  {\thmname{#1}\thmnumber{ #2}\ifx#3\empty\else\ (#3)\fi}
\theoremstyle{definition}

\newtheorem{remark}[theorem]{\underline{Remark}}

\usepackage{sectsty}
\allsectionsfont{\mdseries\scshape}

\usepackage{pdfrender}
\newcommand{\heavytext}[1]{\textpdfrender{
  TextRenderingMode=FillStroke,
  LineWidth=.3pt,
}{#1}}

\renewcommand{\)}{\right)}
\renewcommand{\[}{\left[}
\renewcommand{\]}{\right]}

\newcommand\Eb{\mathds{E}}

\newcommand\Pb{\mathds{P}}

\newcommand\Rb{\mathds{R}}

\newcommand\Cc{\mathscr{C}}

\newcommand\Fc{\mathscr{F}}
\newcommand\Gc{\mathscr{G}}

\newcommand\Nc{\mathscr{N}}

\newcommand\Om{\Omega}

\newcommand\lam{\lambda}
\newcommand\gam{\gamma}
\newcommand\Gam{\Gamma}

\renewcommand\d{\partial}

\newcommand\dd{\mathrm{d}}
\newcommand\ee{\mathrm{e}}

\usepackage{listings}
\definecolor{mma_purple}{rgb}{0.5, 0, 0.5}
\definecolor{mma_blue}{rgb}{0, 0, 1}
\definecolor{mma_green}{rgb}{0, 0.5, 0}
\definecolor{mma_gray}{rgb}{0.5, 0.5, 0.5}
\definecolor{mma_background}{rgb}{0.98, 0.98, 0.98}
\begin{document}

%\title{Optimal Portfolio Management for Yield-Bearing Stablecoins:\\
%A Stochastic Control Approach to the Ethena Protocol}
\title{Optimal Control of the Ethena Yield-Bearing Stablecoin}

\author{
Matthew Lorig
\thanks{Department of Applied Mathematics, University of Washington.
\heavytext{e-mail}: \url{mlorig@uw.edu}.}
}

\date{This version: \today}

\maketitle

\begin{abstract}
We formulate and solve stochastic control problems that model the core
yield-generating strategy of the Ethena protocol, a decentralized finance (DeFi)
stablecoin that earns yield by combining a long position in staked Ethereum (stETH)
with an equal-sized short position in ETH perpetual futures.
The combined position is delta-neutral with respect to the ETH spot price, yet earns
carry from two sources: staking rewards on the stETH leg, and funding-rate payments
received from long perpetual holders when the perpetual trades at a premium to spot.
A key feature of our model is that the control --- the rate of simultaneously buying
stETH and shorting the perpetual --- exerts two distinct types of price impact.
\textit{Permanent} impact shifts the mid-market prices of both legs, compressing the
basis and permanently eroding future funding income; this is modeled with separate
spot and perpetual permanent-impact coefficients $\mu_1$ and $\mu_2$.
\textit{Temporary} impact reflects execution slippage on each leg, modeled with
separate coefficients $\lam_1$ and $\lam_2$; it enters only as a direct
quadratic execution cost $\lam\,\gam_t^2$ with $\lam = \lam_1+\lam_2$.
We study both an infinite-horizon discounted problem and a finite-horizon problem
in which the protocol maximizes total wealth up to a fixed date $T$, subject to
a terminal cost for liquidating any remaining position.
In both cases the optimal control is obtained explicitly: in the infinite-horizon
case via a quadratic ansatz for the Hamilton--Jacobi--Bellman (HJB) equation, and
in the finite-horizon case via a time-dependent quadratic ansatz that reduces the
HJB partial differential equation to a system of Riccati ordinary differential
equations.
In both settings the optimal control is linear in the state $(D_t^\gam, N_t^\gam)$:
the protocol should build its position faster when the basis is wide and taper
accumulation as the position grows and inventory risk accumulates.
Near the terminal date the finite-horizon policy additionally generates a liquidation
urgency that is absent from the infinite-horizon solution.
\end{abstract}

\noindent \heavytext{Keywords}: stochastic control, Hamilton--Jacobi--Bellman equation,
perpetual futures, funding rate, staking yield, price impact, DeFi, stablecoin,
Ornstein--Uhlenbeck process.

%-----------------------------------------------------------------------------------
%
%       SECTION 1:    Introduction
%
%-----------------------------------------------------------------------------------

\section{Introduction}
\label{sec:introduction}

Decentralized finance (DeFi) has given rise to a new class of yield-bearing stablecoins
that seek to combine price stability with attractive returns through explicitly
structured financial strategies.
Among the most prominent of these is the Ethena protocol \citep{ethena2024}, which
issues a dollar-denominated stablecoin whose yield is generated by a
\textit{delta-neutral} strategy: deposits of USD-denominated collateral are used to
purchase staked Ether (stETH), while an equal notional short position in ETH perpetual
futures is simultaneously established.
Because the long stETH position and the short perpetual position respond identically
to movements in the ETH spot price, the combined book is market-neutral in ETH, yet
earns two streams of carry.
First, the stETH earns a continuous staking yield.
Second, because the ETH perpetual futures price typically trades at a premium to the
spot price, the funding mechanism of perpetual contracts requires long holders to make
periodic cash payments to short holders; as the short leg of the trade, Ethena receives
these payments.
\\[0.5em]
The Ethena protocol has grown remarkably quickly since USDe launched in
February 2024 \citep{messari2024ethena}.
Within four months of launch, USDe surpassed a \$3 billion market
capitalization to become the fourth-largest stablecoin, a milestone its
founder claimed had been reached faster than any other USD-denominated
crypto asset in history \citep{binance2024usde}.
Growth accelerated further in 2025: by August of that year USDe had
climbed to a market capitalization of approximately \$9.3 billion,
briefly ranking it as the third-largest stablecoin behind only Tether's
USDT and Circle's USDC \citep{theblock2025usde}, before reaching a peak
of nearly \$15 billion later in the year \citep{coingecko2025annual}.
As of the time of writing (May 2026), USDe carries a market
capitalization of approximately \$4 billion, ranking it among the
thirtieth-largest cryptocurrencies overall and making it the
largest stablecoin that is not backed by government bonds or fiat
bank deposits \citep{coingecko2026usde, coinmarketcap2026usde}.
The protocol's rapid ascent has attracted considerable attention from
both practitioners and academics, and raises fundamental questions about
the optimal dynamic management of the delta-neutral strategy that
underlies it.
\\[0.5em]
The funding-rate mechanism in perpetual futures markets is designed to keep the
perpetual price anchored to the underlying spot.
When the perpetual trades at a premium (positive basis), longs pay shorts at a rate
proportional to the size of the premium; see \citet{bitmex2016funding}.
Empirically, ETH perpetual markets have spent the majority of historical time in
positive-funding environments, reflecting persistent net demand for leveraged long
exposure \citep{deribit2022perps}, making the short perpetual overlay structurally
advantageous.
\\[0.5em]
While the strategy is conceptually simple, its \textit{optimal dynamic implementation}
is non-trivial and involves several interacting forces.
When the protocol grows its position --- buying stETH on the spot market and shorting
the perpetual --- it exerts two distinct types of price impact.
First, there is \textit{permanent} impact: buying stETH shifts the spot mid-price
upward, and shorting the perpetual shifts the perpetual mid-price downward; both
effects compress the perpetual-spot basis and permanently erode the funding premium
from which the protocol earns its carry.
Second, there is \textit{temporary} impact: market orders on each leg fill at prices
worse than the mid, generating immediate execution slippage.
We model these two channels with entirely separate parameters, and show that both
must be controlled in an optimal implementation.
At the same time, a large outstanding position creates inventory risk: in episodes
of negative funding (when the basis inverts and longs receive payments from shorts),
the protocol incurs carry costs proportional to its short position size.
\\[0.5em]
The literature on optimal execution and inventory management provides the natural
framework for these trade-offs.
\citet{almgren2001optimal} pioneered the use of quadratic execution costs to model
market impact in optimal liquidation, yielding tractable linear-quadratic stochastic
control problems with explicit solutions.
\citet{cartea2015algorithmic} and \citet{gueant2016financial} provide comprehensive
treatments of optimal trading with stochastic price dynamics and inventory penalties.
Mean-reverting spread processes similar to our basis model appear in pairs trading and
statistical arbitrage; see \citet{jurek2007dynamic} and \citet{liu2005dynamic}.
Stochastic control in DeFi has received growing attention: \citet{cartea2023decentralised}
study optimal liquidity provision in automated market makers, and perpetual futures
mechanics are analyzed in \citet{ackerer2024perpetuals}.
\\[0.5em]
Our contributions are as follows.
We formulate continuous-time stochastic control problems capturing the Ethena
trade-offs: permanent two-sided price impact from simultaneous spot and perpetual
order flow (which feeds back into the basis dynamics), separate temporary execution
costs from market-order slippage on each leg, carry income from staking and funding,
and a quadratic inventory risk penalty.
We treat both an infinite-horizon discounted formulation and a finite-horizon
formulation in which the protocol maximizes total accumulated wealth up to a fixed
liquidation date.
In both cases we obtain the value function and optimal control explicitly.
In the infinite-horizon case the solution follows from a quadratic ansatz for the
stationary HJB equation, reducing the problem to a cascade of scalar algebraic
equations.
In the finite-horizon case the same quadratic ansatz reduces the HJB PDE to a
backward system of Riccati ODEs, which inherits the same hierarchical structure and
admits an explicit integral formula for the affine (non-quadratic) coefficients.
In both settings the optimal control is a linear feedback in the current basis and
position size; near the terminal date the finite-horizon feedback coefficients generate a liquidation
urgency absent from the infinite-horizon solution.
\\[0.5em]
The rest of this paper is organized as follows.
Section~\ref{sec:model} presents the model.
Section~\ref{sec:IH} formulates and solves an infinite-horizon stochastic control problem.
%treats the infinite-horizon problem: value function, HJB equation, explicit solution, and economic implications.
Section~\ref{sec:FH} formulates and solves a finite-horizon stochastic control problem.
%treats the finite-horizon problem: value function, HJB PDE,
%explicit Riccati solution, and economic implications.
Some numerical experiments are performed in Section~\ref{sec:numerical}.
Section \ref{sec:conclusion} offers some concluding remarks and possible directions for future research.

%-----------------------------------------------------------------------------------
%
%       SECTION 2:    Model and Problem Formulation
%
%-----------------------------------------------------------------------------------

\section{Model}
\label{sec:model}

\subsection{Positions, Prices, and Execution Costs}
\label{sec:dynamics}

Let $(\Om, \Fc, \{\Fc_t\}_{t \geq 0}, \Pb)$ be a filtered probability space satisfying
the usual conditions, and let $W = \{W_t\}_{t \geq 0}$ be a standard Brownian motion.
The Ethena protocol simultaneously holds a long position of $N_t^\gam$ units of staked ETH
and a short position of $N_t^\gam$ units of the ETH perpetual futures contract.
The position size $N_t^\gam \in \Rb$ can be positive (the intended long-stETH, short-perp
carry trade) or negative (a reversed position); as we show in
Section~\ref{sec:IH}, the optimal policy generically keeps $N^\gam > 0$.
%when $m > 0$ and $r > 0$.
The control variable is $\gam_t = \dd N_t^\gam/\dd t$: the rate of buying stETH and
simultaneously shorting the perpetual when $\gam_t > 0$, or unwinding both legs when $\gam_t < 0$.
Here and throughout this paper, stochastic processes that depend on $\gam$ will carry it as a superscript.
\\[0.5em]
%Let $(\Om, \Fc, \{\Fc_t\}_{t \geq 0}, \Pb)$ be a filtered probability space satisfying
%the usual conditions, and let $W = \{W_t\}_{t \geq 0}$ be a standard Brownian motion.
%The Ethena protocol dynamically manages its position at rate $\gam_t \in \Rb$: when
%$\gam_t > 0$ the protocol buys stETH and simultaneously shorts the perpetual; when
%$\gam_t < 0$ it unwinds both legs.
%Throughout the paper, any stochastic process whose evolution depends on the choice of
%control $\gam = \{\gam_t\}_{t \geq 0}$ will carry a superscript $\gam$ to make this
%dependence explicit.
%The protocol simultaneously holds a long position of $N_t^\gam$ units of staked ETH
%and a short position of $N_t^\gam$ units of the ETH perpetual futures contract, where
%the common position size $N_t^\gam \in \Rb$ evolves in response to the control via
%$\dd N_t^\gam = \gam_t \, \dd t$.
%The position size can be positive (the intended long-stETH, short-perp carry trade)
%or negative (a reversed position); as we show in Section~\ref{sec:IH}, the optimal
%policy generically keeps $N^\gam > 0$.
%\\[0.5em]
Let $S_t^\gam$ denote the mid-market spot price of ETH and $P_t^\gam$ the mid-market price of
the ETH perpetual futures, so that the \textit{basis} $D_t^\gam := P_t^\gam - S_t^\gam$ is the
perpetual premium over spot.
When $D^\gam > 0$, the perpetual trades at a premium and the funding mechanism pays longs
to shorts; the Ethena protocol, being short the perpetual, receives these payments.
\\[0.5em]
When the Ethena protocol trades at rate $\gam_t$, it faces two-sided \textit{temporary} execution
costs reflecting the fact that market orders fill at prices worse than the mid.
On the spot leg, buying stETH at rate $\gam_t$ moves the execution price above the
mid by $\lam_1 \gam_t$, so the protocol pays $S_t^\gam + \lam_1\,\gam_t$ per unit purchased,
where $\lam_1 \geq 0$ is the spot temporary impact coefficient.
On the perpetual leg, shorting at rate $\gam_t$ moves the execution price below the mid
by $\lam_2 \gam_t$, so the protocol receives $P_t^\gam - \lam_2\,\gam_t$ per unit shorted,
where $\lam_2 \geq 0$ is the perpetual temporary impact coefficient.
The overpayment on the spot leg is $\lam_1 \gam_t$ per unit times $\gam_t$ units per unit time,
contributing a cost of $\lam_1 \gam_t^2$ per unit time; similarly the perpetual leg
contributes $\lam_2 \gam_t^2$.
The total instantaneous temporary execution cost is therefore
\begin{align}
\text{temporary execution cost rate} = (\lam_1 + \lam_2)\,\gam_t^2
    =: \lam\,\gam_t^2,
\label{eq:execcost}
\end{align}
where we have defined $\lam := \lam_1 + \lam_2 > 0$.

\subsection{Basis Dynamics and Permanent Price Impact}
\label{sec:basis}

Separately from the temporary execution costs above, trading at rate $\gam_t$ also has a
\textit{permanent} effect on mid-market prices that persists after the trade is
complete.
Buying stETH at rate $\gam_t$ shifts the spot mid $S^\gam$ upward permanently at rate
$\mu_1 \gam_t$, and shorting the perpetual at rate $\gam_t$ shifts the perpetual mid $P^\gam$
downward permanently at rate $\mu_2 \gam_t$, where $\mu_1, \mu_2 \geq 0$ are the
permanent impact coefficients (distinct from the temporary impact coefficients $\lam_1,
\lam_2$).
As $D^\gam = P^\gam - S^\gam$, both permanent shifts act in the same direction to compress the
basis:
\begin{align}
\text{permanent impact on }D^\gam:
\quad -(\mu_1 + \mu_2)\,\gam_t\,\dd t
    =: -\mu\, \gam_t \dd t, \qquad \mu := \mu_1 + \mu_2.
\end{align}
Absent order flow, the basis mean-reverts to a long-run level $m > 0$ at speed
$\kappa > 0$ and is subject to random shocks.
Combining the natural mean-reversion, the permanent impact, and the noise, the
controlled basis dynamics are
\begin{align}
\dd D_t^\gam
    = \Big[-\kappa\(D_t^\gam - m\) - \mu\,\gam_t\Big]\dd t + c\,\dd W_t,
\label{eq:dD}
\end{align}
where $c > 0$ is the basis volatility.
Because the impact is permanent, the shift $-\mu \gam_t\,\dd t$ alters $D_t^\gam$ for all future
times: the protocol must internalize that aggressive accumulation today permanently
erodes the funding premium it will earn tomorrow.
The position size evolves as
\begin{align}
\dd N_t^\gam = \gam_t\,\dd t, 
%\qquad N_t^\gam \in \Rb.
\label{eq:dN}
\end{align}
The state of the system is the pair $(D_t^\gam, N_t^\gam) \in \Rb^2$.

\subsection{Net Wealth Dynamics}
\label{sec:wealth}

Let $X_t^\gam$ denote the net wealth of the Ethena protocol.  
Changes in the value of $X_t^\gam$ over the period $[t, t+\dd t]$ arise from three sources.

\medskip
\noindent\textit{(i) Carry income.}
The long stETH position earns staking income at constant rate $r > 0$ per unit,
contributing $r\,N_t^\gam\,\dd t$.
The short perpetual position receives funding payments at rate $q\,D_t^\gam$ per unit
(positive when $D > 0$, i.e.\ when longs pay shorts), contributing
$q\,D_t^\gam\,N_t^\gam\,\dd t$.

\medskip
\noindent\textit{(ii) Mark-to-market P\&L.}
The long stETH leg has mark-to-market (MTM) gain $N_t^\gam\,\dd S_t^\gam$ and the short perpetual leg has MTM
gain $-N_t^\gam\,\dd P_t^\gam$.
Their sum is $N_t^\gam(\dd S_t^\gam - \dd P_t^\gam) = -N_t^\gam\,\dd D_t^\gam$: the $\dd S_t^\gam$ exposures cancel
exactly (delta-neutrality), leaving only basis risk.
Substituting \eqref{eq:dD}, we have
\begin{align}
-N_t^\gam\,\dd D_t^\gam
    = \Big[\kappa\(D_t^\gam-m\) + \mu\,\gam_t\Big]N_t^\gam\,\dd t
      - c\,N_t^\gam\,\dd W_t.
\label{eq:mtm}
\end{align}
The term $\mu \gam_tN_t^\gam\,\dd t$ is the immediate MTM gain from the protocol's own
permanent impact: the spot mid rises by $\mu_1 \gam_t\,\dd t$ (gaining $\mu_1 \gam_t N_t^\gam$
on the long stETH leg) and the perpetual mid falls by $\mu_2 \gam_t\,\dd t$ (gaining
$\mu_2 \gam_t N_t^\gam$ on the short perp leg), summing to $\mu \gam_t N_t^\gam\,\dd t$.
Note that this gain is not ``free'': it is 
%exactly 
offset in expectation by the permanently
lower future funding income caused by the compressed basis.

\medskip
\noindent\textit{(iii) Execution cost.}
The total execution cost derived in \eqref{eq:execcost} is $\lam\,\gam_t^2\,\dd t$.

\medskip
\noindent Combining (i), (ii) and (iii), we have
\begin{align}
\dd X_t^\gam
		&=	r \, N_t^\gam \dd t + q\,D_t^\gam\,N_t^\gam\,\dd t - N_t^\gam\,\dd D_t^\gam - \lam\,\gam_t^2\,\dd t \\
    &= \Big[\(q + \kappa\)D_t^\gam - \kappa m + r + \mu\,\gam_t\Big]N_t^\gam\,\dd t
      - \lam\,\gam_t^2\,\dd t
      - c\,N_t^\gam\,\dd W_t.
\label{eq:dX}
\end{align}

\begin{remark}[Permanent vs.\ temporary impact]
\label{rem:impact}
The model contains two distinct impact channels with separate parameters.
The \textit{permanent} impact coefficients $\mu_1$ (spot) and $\mu_2$ (perp)
govern how much the mid-market prices shift lastingly per unit of order flow; their
sum $\mu = \mu_1+\mu_2$ enters the basis SDE \eqref{eq:dD} and generates the MTM
gain $\mu \gam_t N_t^\gam\,\dd t$ in \eqref{eq:dX}.
The \textit{temporary} impact coefficients $\lam_1$ (spot) and $\lam_2$ (perp)
determine the execution-price slippage
$\lam = \lam_1+\lam_2$ that appears only as the direct cost $\lam\,\gam_t^2\,\dd t$
in the objective.
The two sets of parameters are independent: $\mu$ and $\lam$ can take any non-negative
values.
%, and in particular there is no constraint $\lam \geq \mu$.
We use the term ``permanent impact'' in the standard Almgren--Chriss sense: each unit
of order flow shifts the mid-price persistently from the time of the trade onward.
Because the basis is also subject to exogenous mean reversion at speed $\kappa$, the
net effect of a past trade on the basis decays over time; ``permanent'' refers to the
absence of an immediate price-reversion component, not to the long-run level of the basis.
\end{remark}

%-----------------------------------------------------------------------------------
%
%       SECTION 3:    Infinite-Horizon Stochastic Control Problem
%
%-----------------------------------------------------------------------------------

\section{Infinite-Horizon Stochastic Control Problem}
\label{sec:IH}

\subsection{Value Function and HJB Equation}
\label{sec:IH-hjb}

We introduce a quadratic inventory risk penalty $\phi \cdot (N_t^\gam)^2$ (with
$\phi > 0$) to penalize large positions, capturing basis variance risk (the term
$-cN_t^\gam\,\dd W_t$ in \eqref{eq:dX} has variance $c^2 \cdot (N_t^\gam)^2\,\dd t$),
liquidation risk, and the cost of sustained negative funding.
Let $\Gc$ denote the class of admissible controls consisting of progressively
measurable processes $\gam = \{\gam_t\}_{t \geq 0}$ for which
\eqref{eq:dD}--\eqref{eq:dN} admit a unique strong solution.
We define the infinite-horizon value function as follows
\begin{align}
V(d,n)
    &:= \sup_{\gam\,\in\,\Gc}\;
       \Eb\[\int_0^\infty \ee^{-\rho t} \Big(
            \dd X_t^\gam - \phi \cdot (N_t^\gam)^2 \dd t
            \Big) \;\Big|\;D_0^\gam=d,\;N_0^\gam=n\] \notag \\
    &= \sup_{\gam\,\in\,\Gc}\;
       \Eb\[\int_0^\infty \ee^{-\rho t}\Big(
           \big[(q+\kappa)D_t^\gam - \kappa m + r + \mu\,\gam_t\big]N_t^\gam
           - \lam\,\gam_t^2
           - \phi \cdot (N_t^\gam)^2
       \Big)\dd t \;\Big|\;D_0^\gam=d,\;N_0^\gam=n\],
\label{eq:V}
\end{align}
where $\rho > 0$ is the discount rate.
The three terms in the integrand represent, respectively: the expected rate of wealth
increase from carry income and permanent-impact MTM; the total execution cost; and the
inventory risk penalty.
\\[0.5em]
By the principle of dynamic programming, the value function $V:\Rb^2\to\Rb$ satisfies
the HJB equation
\begin{align}
\rho\,V(d,n)
    = \sup_{\gam\,\in\,\Rb}\Big\{
        \big[(q+\kappa)d - \kappa m + r + \mu \gam\big]n
        - \lam \gam^2 - \phi n^2
        + \big[-\kappa(d-m) - \mu \gam\big]\d_d V
        + \gam\,\d_n V
        + \tfrac{c^2}{2}\,\d_{dd}V
    \Big\}.
\label{eq:HJB}
\end{align}
The supremum over $\gam$ is attained at the unconstrained maximizer obtained from the
first-order condition
\begin{align}
\mu n - 2\lam \gam - \mu\,\d_d V + \d_n V = 0,
\label{eq:FOC}
\end{align}
giving the candidate optimal feedback control
\begin{align}
\gam^*(d,n) = \frac{\mu\,n - \mu\,\d_d V(d,n) + \d_n V(d,n)}{2\lam}.
\label{eq:gstar-general}
\end{align}
The term $\mu n$ in the numerator of \eqref{eq:gstar-general} --- absent in standard
execution problems with only temporary impact --- arises from the permanent impact MTM
benefit $\mu \gam N^\gam$ in the running reward: a larger existing position makes it more
attractive to keep accumulating, since each new unit sold short generates an immediate
MTM gain of $\mu \gam\,\dd t$ per existing unit.
Substituting \eqref{eq:gstar-general} into \eqref{eq:HJB} gives the reduced HJB:
\begin{align}
\rho\,V
    = \big[(q+\kappa)d - \kappa m + r\big]n - \phi n^2
      - \kappa(d-m)\,\d_d V
      + \frac{\big(\mu n - \mu\,\d_d V + \d_n V\big)^2}{4\lam}
      + \frac{c^2}{2}\,\d_{dd}V.
\label{eq:HJB-reduced}
\end{align}

\subsection{Explicit Solution}
\label{sec:IH-solution}

The linear-quadratic structure of \eqref{eq:HJB-reduced} motivates the ansatz
\begin{align}
V(d,n) = \alpha_1 n^2 + \alpha_2 nd + \alpha_3 n + \alpha_4 d^2 + \alpha_5 d + \alpha_6,
\label{eq:ansatz}
\end{align}
with coefficients $\alpha_1,\ldots,\alpha_6\in\Rb$ to be determined.
The partial derivatives are
\begin{align}
\d_d V = \alpha_2 n + 2\alpha_4 d + \alpha_5,
\qquad
\d_n V = 2\alpha_1 n + \alpha_2 d + \alpha_3,
\qquad
\d_{dd}V = 2\alpha_4.
\label{eq:partials}
\end{align}
Define the combined quantity
\begin{align}
\Phi(d,n)
    &:= \mu n - \mu\,\d_d V + \d_n V \notag\\
    &= \big(\mu(1 - \alpha_2) + 2\alpha_1\big)n
       + \big(\alpha_2 - 2\mu\alpha_4\big)d
       + \big(\alpha_3 - \mu\alpha_5\big).
\label{eq:Phi}
\end{align}
If the value function is of the form \eqref{eq:ansatz}, we can express the optimal control as
%In terms of the feedback gains defined below, 
$\gam^* = \Phi/(2\lam) = \gam_N n + \gam_D d + \gam_0$, where
\begin{align}
%\gam_N &:= \frac{\mu(1-\alpha_2) + 2\alpha_1}{2\lam},
%\label{eq:gamN}\\
%\gam_D &:= \frac{\alpha_2 - 2\mu\alpha_4}{2\lam},
%\label{eq:gamD}\\
%\gam_0 &:= \frac{\alpha_3 - \mu\alpha_5}{2\lam}.
%\label{eq:gam0}
\gam_N &:= \frac{\mu(1-\alpha_2) + 2\alpha_1}{2\lam}, &
\gam_D &:= \frac{\alpha_2 - 2\mu\alpha_4}{2\lam}, &
\gam_0 &:= \frac{\alpha_3 - \mu\alpha_5}{2\lam}. &
\label{eq:gamN}
\end{align}
We are now in a position to present our main result for the infinite-horizon stochastic control problem.

%\subsubsection{Main Result}
%\label{sec:IH-main}

%\begin{proposition}[Explicit Solution]
%\label{prop:main}
%Suppose $\rho > 0$, $\kappa > 0$, $\phi > 0$, $\mu = \mu_1+\mu_2 \geq 0$,
%$\lam = \lam_1+\lam_2 > 0$, $r > 0$, $q > 0$, and the stability condition
%\begin{align}
%\kappa_* := \rho + 2\kappa - c^2 > 0.
%\label{eq:stability}
%\end{align}
%Then the value function \eqref{eq:V} is given by the quadratic \eqref{eq:ansatz} and
%the optimal control is the linear feedback
%\begin{align}
%\gam^*(D_t^\gam,N_t^\gam) = \gam_N\,N_t^\gam + \gam_D\,D_t^\gam + \gam_0,
%\label{eq:gstar}
%\end{align}
%where the feedback coefficients $\gam_N$, $\gam_D$, $\gam_0$ are given by
%\eqref{eq:gamN}--\eqref{eq:gam0} and the coefficients $\alpha_1,\ldots,\alpha_6$ are
%determined by the explicit cascade in the proof below.
%\end{proposition}

\begin{proposition}[Explicit Solution]
\label{prop:main}
Suppose $\rho > 0$, $\kappa > 0$, $\phi > 0$, $\mu = \mu_1+\mu_2 > 0$,
$\lam = \lam_1+\lam_2 > 0$, $r > 0$, $q > 0$, and the condition
\begin{align}
\kappa_* := \rho + 2\kappa > 0.
\label{eq:stability}
\end{align}
Then the value function \eqref{eq:V} is given by the quadratic \eqref{eq:ansatz} and
the optimal control is the linear feedback
\begin{align}
\gam^*(D_t^\gam,N_t^\gam) = \gam_N\,N_t^\gam + \gam_D\,D_t^\gam + \gam_0,
\label{eq:gstar}
\end{align}
where the feedback coefficients $\gam_N$, $\gam_D$, $\gam_0$ are given by
%\eqref{eq:gamN}--\eqref{eq:gam0}, 
\eqref{eq:gamN}, 
the coefficients $\alpha_1$ and $\alpha_4$ are
the admissible roots of the quadratics \eqref{eq:alpha1-quadratic} and
\eqref{eq:alpha4-quadratic}, given explicitly by \eqref{eq:alpha1-sol} and
\eqref{eq:alpha4-sol}, the coefficient $\alpha_2$ solves the fixed-point equation
\eqref{eq:sys2}, the coefficients $\alpha_3$ and $\alpha_5$ solve the linear system
\eqref{eq:lin1}--\eqref{eq:lin2} with explicit solution \eqref{eq:C-sol}, and
$\alpha_6$ is given by \eqref{eq:sys6}.
\end{proposition}

\begin{proof}
Substitute \eqref{eq:ansatz}--\eqref{eq:partials} into the reduced HJB
\eqref{eq:HJB-reduced}, expand $\Phi^2$, and collect by monomial in $(d,n)$.

\medskip
\noindent\textit{Coefficient of $n^2$:}
\begin{align}
\rho\alpha_1 = -\phi + \lam\gam_N^2,
\qquad\text{i.e.,}\qquad
4\lam\rho\alpha_1
    = -4\lam\phi + \big[\mu(1-\alpha_2)+2\alpha_1\big]^2.
\label{eq:sys1}
\end{align}
Expanding and collecting terms in $\alpha_1$:
\begin{align}
4\alpha_1^2
    + \big[4\mu(1-\alpha_2) - 4\lam\rho\big]\alpha_1
    + \mu^2(1-\alpha_2)^2 - 4\lam\phi = 0.
\label{eq:alpha1-quadratic}
\end{align}
The admissible root (giving $\alpha_1 < 0$ so that $V$ is concave in $n$) is
\begin{align}
\alpha_1
    = \frac{\big[\lam\rho - \mu(1-\alpha_2)\big]
            - \sqrt{\big[\lam\rho - \mu(1-\alpha_2)\big]^2
                   + 4\lam\phi - \mu^2(1-\alpha_2)^2}}{2}.
\label{eq:alpha1-sol}
\end{align}

\medskip
\noindent\textit{Coefficient of $d^2$:}
\begin{align}
\rho\alpha_4 = \lam\gam_D^2 - 2\kappa\alpha_4,
\qquad\text{i.e.,}\qquad
4\lam\kappa_*\alpha_4 = (\alpha_2 - 2\mu\alpha_4)^2,
\label{eq:sys3}
\end{align}
where $\kappa_* = \rho+2\kappa>0$.
This yields the quadratic
\begin{align}
4\mu^2\alpha_4^2 - \big(4\mu\alpha_2 + 4\lam\kappa_*\big)\alpha_4 + \alpha_2^2 = 0,
\label{eq:alpha4-quadratic}
\end{align}
with admissible root
\begin{align}
\alpha_4
    = \frac{\big(\mu\alpha_2 + \lam\kappa_*\big)
            - \sqrt{\big(\mu\alpha_2 + \lam\kappa_*\big)^2 - \mu^2\alpha_2^2}}
           {2\mu^2}
    = \frac{\big(\mu\alpha_2 + \lam\kappa_*\big)
            - \sqrt{\lam\kappa_*\big(\lam\kappa_* + 2\mu\alpha_2\big)}}
           {2\mu^2}.
\label{eq:alpha4-sol}
\end{align}

\medskip
\noindent\textit{Coefficient of $nd$:}
\begin{align}
(\rho+\kappa)\alpha_2
    = (q+\kappa)
      + 2\lam\gam_N\gam_D
    = (q+\kappa)
      + \frac{\big[\mu(1-\alpha_2)+2\alpha_1\big]\big(\alpha_2-2\mu\alpha_4\big)}{2\lam}.
\label{eq:sys2}
\end{align}
Since $\alpha_1$ and $\alpha_4$ depend on $\alpha_2$ through
\eqref{eq:alpha1-sol}--\eqref{eq:alpha4-sol}, equation \eqref{eq:sys2} is a scalar
nonlinear fixed-point equation for $\alpha_2$.
It can be solved by iteration; see Remark~\ref{rem:roots} for the choice of starting point.

\medskip
\noindent\textit{Coefficients of $n$ and $d$ (linear terms):}
\noindent
Collecting the coefficient of $n$ (constant in $d$), using $-\kappa(d-m)\d_d V \ni
\kappa m\alpha_2 n$ and the cross term $\Phi^2/(4\lam) \ni \gam_N\gam_0 n$:
\begin{align}
\rho\alpha_3
    = (r - \kappa m) + \kappa m\alpha_2
      + \frac{\big[\mu(1-\alpha_2)+2\alpha_1\big]\big(\alpha_3-\mu\alpha_5\big)}{2\lam}
    = (r - \kappa m) + \kappa m\alpha_2 + \gam_N(\alpha_3 - \mu\alpha_5).
\label{eq:sys4}
\end{align}
Collecting the coefficient of $d$ (constant in $n$), using $-\kappa(d-m)\d_d V \ni
-\kappa\alpha_5 d + 2\kappa m\alpha_4 d$ and the cross term $\ni \gam_D\gam_0 d$:
\begin{align}
(\rho+\kappa)\alpha_5
    = 2\kappa m\alpha_4
      + \frac{\big(\alpha_2-2\mu\alpha_4\big)\big(\alpha_3-\mu\alpha_5\big)}{2\lam}
    = 2\kappa m\alpha_4 + \gam_D(\alpha_3 - \mu\alpha_5).
\label{eq:sys5}
\end{align}
Writing $C := \alpha_3 - \mu\alpha_5$, equations \eqref{eq:sys4}--\eqref{eq:sys5}
become the $2\times 2$ linear system
\begin{align}
(\rho - \gam_N)\alpha_3 + \gam_N\mu\,\alpha_5 &= (r - \kappa m) + \kappa m\alpha_2, \label{eq:lin1}\\
-\gam_D\alpha_3 + (\rho+\kappa+\gam_D\mu)\alpha_5 &= 2\kappa m\alpha_4. \label{eq:lin2}
\end{align}
Setting $R := (r-\kappa m)+\kappa m\alpha_2$ and $S := 2\kappa m\alpha_4$, the
determinant of the coefficient matrix is
\begin{align}
\Delta
    := (\rho-\gam_N)(\rho+\kappa+\gam_D\mu) + \gam_N\mu\gam_D
    = (\rho-\gam_N)(\rho+\kappa) + \rho\mu\gam_D.
\label{eq:Delta}
\end{align}
Applying Cramer's rule and using $C = \alpha_3 - \mu\alpha_5$:
\begin{align}
C
    = \frac{R(\rho+\kappa) - \rho\mu S}{\Delta}
    = \frac{(r-\kappa m)(\rho+\kappa) + \kappa m\alpha_2(\rho+\kappa) - 2\rho\mu\kappa m\alpha_4}
           {(\rho-\gam_N)(\rho+\kappa) + \rho\mu\gam_D},
\label{eq:C-sol}
\end{align}
and then $\alpha_3 = (R + \gam_N C)/\rho$, $\alpha_5 = (\alpha_3 - C)/\mu$.

\medskip
\noindent\textit{Constant term:}
\begin{align}
\rho\alpha_6 = \lam\gam_0^2 + c^2\alpha_4 + \kappa m\alpha_5.
\label{eq:sys6}
\end{align}
This completes the determination of all six coefficients.
\end{proof}

\begin{remark}[Admissibility conditions and the case $\mu=0$]
\label{rem:admissibility}
Proposition~\ref{prop:main} is stated conditionally: the displayed formulas give the
candidate solution provided (i) the discriminant in \eqref{eq:alpha1-sol} is
non-negative, i.e., $4\lam\phi \geq \mu^2(1-\alpha_2)^2$; (ii) the discriminant in
\eqref{eq:alpha4-sol} is non-negative, i.e., $\lam\kappa_*(\lam\kappa_*+2\mu\alpha_2)\geq 0$;
(iii) the fixed-point equation \eqref{eq:sys2} has an admissible solution $\alpha_2$;
and (iv) the determinant $\Delta \neq 0$ so that the linear system
\eqref{eq:lin1}--\eqref{eq:lin2} has a unique solution.
These conditions hold for the parameters of Section~\ref{sec:numerical} and can be verified numerically before use.
\\[0.5em]
The restriction $\mu > 0$ in the proposition is necessary because
\eqref{eq:alpha4-sol} divides by $\mu^2$ and the formula $\alpha_5 = (\alpha_3-C)/\mu$
divides by $\mu$.
When $\mu=0$, \eqref{eq:sys3} gives $\alpha_4 = \alpha_2^2/(4\lam\kappa_*)$ directly,
and the two linear equations \eqref{eq:sys4}--\eqref{eq:sys5} decouple:
$\alpha_3 = R/(\rho - \gam_N)$ and $\alpha_5 = (S + \gam_D\alpha_3)/(\rho+\kappa)$,
where $R = (r-\kappa m)+\kappa m\alpha_2$ and $S = 2\kappa m\alpha_4$,
provided $\rho \neq \gam_N$.
\end{remark}

\begin{remark}[Root selection and concavity]
\label{rem:roots}
The minus sign before the square root in \eqref{eq:alpha1-sol} yields $\alpha_1 < 0$,
ensuring that $V$ is strictly concave in $n$ and the supremum in \eqref{eq:V} is
finite.
Concavity in $n$ requires $4\lam\phi > \mu^2(1-\alpha_2)^2$; when $\phi$ is large
relative to $\mu^2/\lam$ this is readily satisfied.
The root \eqref{eq:alpha4-sol} with the minus sign gives $\alpha_4 \geq 0$: the value
function is convex in $d$, reflecting the fact that a wider basis is unambiguously
beneficial (more funding income).
Standard verification arguments \citep{fleming2006controlled} confirm the candidate
solution equals the true value function.
When solving \eqref{eq:sys2} numerically, a natural starting guess is
$\alpha_2^{(0)} = (q+\kappa)/(\rho+\kappa)$ (the solution when the coupling terms on
the right vanish), but this may violate the concavity condition
$4\lam\phi > \mu^2(1-\alpha_2^{(0)})^2$ for the paper's parameters, rendering the
formula \eqref{eq:alpha1-sol} inadmissible at the initial iterate.
A safe alternative starting point is $\alpha_2^{(0)} = 0$, which satisfies the
concavity condition whenever $4\lam\phi > \mu^2$, as holds for the parameters in
Section~\ref{sec:numerical}.
\end{remark}

\begin{remark}[Closed-loop dynamics]
\label{rem:closed-loop}
Under the optimal control \eqref{eq:gstar}, the joint dynamics of $(D, N)$ are
\begin{align}
\dd D_t^\gam &= \big[-(\kappa + \mu\gam_D)D_t^\gam
              - \mu\gam_N N_t^\gam
              + \kappa m - \mu\gam_0\big]\dd t + c\,\dd W_t,
\label{eq:dD-closed}\\
\dd N_t^\gam &= \big[\gam_N N_t^\gam + \gam_D D_t^\gam + \gam_0\big]\dd t.
\label{eq:dN-closed}
\end{align}
The effective mean-reversion speed of the basis under the optimal policy is
$\kappa + \mu\gam_D \geq \kappa$, which is \emph{faster} than the exogenous
mean-reversion $\kappa$: the protocol's accumulation further compresses the basis,
supplementing the natural reversion.
Stability of the two-dimensional affine system requires the drift matrix
\begin{align}
M :=
\begin{pmatrix}
-(\kappa + \mu\gam_D) & -\mu\gam_N \\
\gam_D & \gam_N
\end{pmatrix}
\label{eq:drift-matrix}
\end{align}
to have eigenvalues with strictly negative real parts.
For a real $2\times 2$ matrix the Hurwitz conditions are equivalent to
$\operatorname{tr}M < 0$ and $\det M > 0$, i.e.,
\begin{align}
\gam_N < \kappa + \mu\gam_D
\qquad\text{and}\qquad
\gam_N < 0.
\label{eq:stability-cl}
\end{align}
These conditions are not implied by the assumptions of Proposition~\ref{prop:main}
alone and should be verified for the chosen parameter values.
For the parameters of Section~\ref{sec:numerical} the eigenvalues of $M$ are
approximately $-5.10$ and $-0.87$, confirming stability in that case.
When \eqref{eq:stability-cl} holds, the stationary distribution of $(D^\gam, N^\gam)$
under \eqref{eq:dD-closed}--\eqref{eq:dN-closed} is Gaussian;
its mean vector and covariance are determined by the linear system
\eqref{eq:mean-ode}--\eqref{eq:lyapunov} evaluated at the stationary limit.
%confirming Remark~\ref{rem:sign-N}.
Since $(D^\gam_t, N^\gam_t)$ is a Gaussian process (being the solution of a linear SDE
with constant diffusion coefficient), $N_t^\gam$ is normally distributed at each fixed
$t$. Writing $m_D(t) := \Eb[D_t^\gam]$, $m_N(t) := \Eb[N_t^\gam]$, and letting
$\Sigma(t)$ be the $2\times 2$ covariance matrix of $(D_t^\gam, N_t^\gam)^\top$,
these satisfy the scalar mean ODEs and Lyapunov equation
\begin{align}
\dot{m}_D(t) &= -(\kappa+\mu\gam_D)\,m_D(t) - \mu\gam_N\,m_N(t) + \kappa m - \mu\gam_0, &
  m_D(0) &= D_0,
\label{eq:mean-ode}\\
\dot{m}_N(t) &= \gam_N\,m_N(t) + \gam_D\,m_D(t) + \gam_0, &
  m_N(0) &= N_0,\\
\dot{\Sigma}(t) &= M\,\Sigma(t) + \Sigma(t)\,M^\top + Q, &
\Sigma(0) &= \mathbf{0},
\label{eq:lyapunov}
\end{align}
where $M$ is the drift matrix \eqref{eq:drift-matrix} and
$Q = \mathrm{diag}(c^2, 0)$.
The probability that $N_t^\gam$ is negative at a fixed time $t$ is
\begin{align}
\Pb(N_t^\gam < 0)
  = \Nc\!\!\left(\frac{-m_N(t)}{\sqrt{\Sigma_{22}(t)}}\right),
\label{eq:PNneg}
\end{align}
where $\Nc(x) = \frac{1}{\sqrt{2\pi}}\int_{-\infty}^x e^{-z^2/2}\,\dd z$ is the
standard normal CDF.
For the probability that $N_t^\gam$ remains non-negative over the entire interval
$[0,T]$, i.e.\ $\Pb(\inf_{t\in[0,T]} N_t^\gam \geq 0)$, a closed-form expression is
not available in general, since $N_t^\gam$ is a Gaussian process (not a Brownian
motion) and computing the distribution of its infimum requires evaluating an
infinite-dimensional integral. This probability can, however, be estimated numerically
from the ODEs \eqref{eq:mean-ode}--\eqref{eq:lyapunov} together with a Monte Carlo
simulation of the closed-loop paths.
Although the current Ethena implementation only reduces its position toward zero,
a reversed position ($N_t^\gam < 0$: short stETH, long perpetual) is economically
coherent and would be optimal when funding rates are persistently negative.
Our model therefore allows $N_t^\gam \in \Rb$, treating the unconstrained problem
in the spirit of the Almgren--Chriss framework and its descendants, where
Gaussian asset-price dynamics routinely produce negative prices without invalidating
the qualitative conclusions of the analysis.
\end{remark}

\subsection{Economic Implications}
\label{sec:IH-discussion}

The closed-form solution of Proposition~\ref{prop:main} reveals several economically
important features of the optimal Ethena strategy.

\paragraph{Funding-rate chasing.}
The feedback coefficient $\gam_D = (\alpha_2 - 2\mu\alpha_4)/(2\lam)$ governs how
aggressively the protocol responds to the current basis $D_t^\gam$.
A wider basis raises both the instantaneous funding income $(q+\kappa)D\cdot N$ and
the mark-to-market benefit of accumulation ($\mu \gam N$), so we expect $\gam_D > 0$:
the protocol should build its position faster when the perpetual premium is high.
Conversely, when the basis is narrow or negative, the optimal policy reduces $\gam_t$
(slows accumulation or unwinds), since funding income is low and the risk of further
basis compression is not worth bearing.

\paragraph{Two-sided permanent impact and basis self-compression.}
A central novelty of our model is the two-sided permanent impact: buying stETH shifts
the spot mid upward at rate $\mu_1 \gam_t$ and shorting the perpetual shifts the
perpetual mid downward at rate $\mu_2 \gam_t$, jointly compressing the basis through
$\mu = \mu_1 + \mu_2$.
The closed-loop basis dynamics \eqref{eq:dD-closed} reveal that the effective
mean-reversion speed becomes $\kappa + \mu\gam_D$, which exceeds the exogenous
$\kappa$: the protocol's own accumulation supplements natural reversion, limiting the
duration of high-funding environments.
The optimal policy internalizes this self-compression and moderates $\gam_D$ as $\mu$
grows.
The individual values of $\mu_1$ and $\mu_2$ do not affect the optimal control
(only $\mu = \mu_1+\mu_2$ enters the HJB), but they determine the P\&L
attribution between the spot and perp legs.

\paragraph{Inventory risk brake.}
The feedback coefficient $\gam_N = [\mu(1-\alpha_2)+2\alpha_1]/(2\lam)$ governs how the optimal rate
responds to the existing position size.
For sufficiently large inventory penalty $\phi$, the term $2\alpha_1$ dominates and
$\gam_N < 0$: the optimal policy is \textit{self-limiting}, reducing the accumulation
rate as $N$ grows.
This reflects the quadratic risk penalty $\phi N^2$, which makes the marginal value
of adding to the position decline as the book grows.
In the closed-loop system \eqref{eq:dN-closed}, this negative feedback ensures that
$N_t^\gam$ remains bounded in expectation.

\paragraph{Baseline accumulation drift.}
The constant feedback coefficient $\gam_0 = (\alpha_3 - \mu\alpha_5)/(2\lam)$ encodes a baseline
rate of accumulation driven by the staking yield $r$ and the long-run basis mean $m$.
Even when $D^\gam = 0$ and $N^\gam = 0$, the protocol optimally accumulates at rate $\gam_0 > 0$
provided $r > 0$ and $m > 0$, reflecting the unconditional attractiveness of the
delta-neutral carry trade.
Starting from zero, $\gam_0$ governs how quickly the protocol initiates its position.

\paragraph{Parameter sensitivities.}
Four key parameters modulate the aggressiveness of the strategy:
\begin{itemize*}
\item A higher \textit{discount rate} $\rho$ makes the protocol more impatient,
  reducing the value of future carry and leading to smaller $\gam_D$ and $\gam_0$
  and a lower steady-state position.
\item A higher \textit{inventory penalty} $\phi$ tightens the risk constraint,
  driving $\alpha_1$ more negative and reducing the equilibrium position size.
  In the limit $\phi\to\infty$ the optimal position converges to zero; as
  $\phi\to 0$ the protocol accumulates without bound (regulated only by the permanent
  impact mechanism).
\item A higher \textit{permanent impact} $\mu = \mu_1+\mu_2$ has two opposing
  effects: it raises the MTM gain $\mu \gam N^\gam$ in the running reward, but also
  compresses the basis more aggressively via \eqref{eq:dD}, destroying future funding
  income faster.
  The optimal policy internalizes this by moderating $\gam_D$ as $\mu$ grows.
  The individual values of $\mu_1$ and $\mu_2$ do not affect the optimal control
  (only their sum $\mu$ enters the HJB), but they determine the P\&L attribution
  between the spot and perpetual legs.
\item A higher \textit{temporary impact} $\lam = \lam_1+\lam_2$ penalizes fast
  trading directly.
  Since all three feedback coefficients scale as $1/\lam$, a larger $\lam$ uniformly slows the
  optimal accumulation, spreading position changes over a longer horizon.
  As $\lam\to 0$, temporary friction vanishes and the only brakes on accumulation
  are the permanent impact $\mu$ and the inventory penalty $\phi$.
  The coefficients $\lam_1$ and $\lam_2$ individually determine the slippage
  attribution between the spot and perpetual legs.
\end{itemize*}

\paragraph{Negative funding.}
When the basis turns negative ($D^\gam < 0$), the carry term $(q+\kappa)D^\gam N^\gam$ becomes
negative for $N^\gam > 0$, imposing a carry cost.
The feedback coefficient $\gam_D > 0$ then drives $\gam^* = \gam_N N^\gam + \gam_D D^\gam + \gam_0$
downward (since $D^\gam < 0$), causing the protocol to reduce its position.
The speed of de-risking is controlled by $|\gam_D|$: a protocol with high $\phi$
(large $|\gam_N|$ and negative $\gam_N$) and large $\mu$ will unwind rapidly, while a
more patient protocol with small $\rho$ may hold through transient negative-funding
episodes and await mean reversion.

\paragraph{Mean-variance interpretation.}
The quadratic inventory penalty $\phi \cdot (N^\gam)^2$ approximates a mean-variance objective:
setting $\phi = \frac{1}{2}c^2\eta$ for risk-aversion coefficient $\eta$ makes $\phi \cdot (N^\gam)^2 = \frac{1}{2} c^2 \eta \cdot (N^\gam)^2$ proportional to the instantaneous variance of the
wealth increment $-cN^\gam\dd W$ in \eqref{eq:dX}.
Under this identification, the optimal policy \eqref{eq:gstar} is the continuous-time
analog of the mean-variance efficient frontier applied to the carry-versus-basis-risk
trade-off.

%-----------------------------------------------------------------------------------
%
%       SECTION 4:    Finite-Horizon Stochastic Control Problem
%
%-----------------------------------------------------------------------------------

\section{Finite-Horizon Stochastic Control Problem}
\label{sec:FH}

\subsection{Value Function and HJB Equation}
\label{sec:FH-hjb}

We now consider a finite-horizon variant in which the protocol operates over a fixed
interval $[0, T]$ and seeks to maximize expected total wealth accumulated by time $T$.
At the terminal date the protocol must close its remaining position $N_T^\gam$ by
liquidating both legs simultaneously.
Liquidating $N_T^\gam$ units at once incurs a lump-sum cost modeled as
$\frac{1}{2}\lam_T (N_T^\gam)^2$, where $\lam_T > 0$ is a terminal liquidation-cost
coefficient reflecting the price impact of closing a large position quickly.
The finite-horizon value function is
\begin{align}
&V^T(t,d,n)
    := \sup_{\gam\,\in\,\Gc}\;
       \Eb\[\int_t^T \Big(
            \dd X_s^\gam - \phi \cdot (N_s^\gam)^2 \dd s
            \Big)
           - \tfrac{1}{2}\lam_T \cdot (N_T^\gam)^2
           \;\Big|\; D_t^\gam = d,\; N_t^\gam = n\] \notag \\
    &= \sup_{\gam\,\in\,\Gc}\;
       \Eb\[\int_t^T\!\Big(
           \big[(q+\kappa)D_s^\gam - \kappa m + r + \mu\,\gam_s\big]N_s^\gam
           - \lam\,\gam_s^2
           - \phi\,(N_s^\gam)^2
       \Big)\dd s
       - \tfrac{1}{2}\lam_T \cdot (N_T^\gam)^2
       \;\Big|\; D_t^\gam = d,\; N_t^\gam = n\],
\label{eq:V-FH}
\end{align}
where we set the discount rate to zero ($\rho = 0$), as is natural when the
objective is total accumulated wealth rather than a discounted present value.
The terminal condition is
\begin{align}
V^T(T, d, n) = -\tfrac{1}{2}\lam_T\,n^2.
\label{eq:terminal}
\end{align}

\begin{remark}[Terminal penalty]
\label{rem:terminal}
The term $-\frac{1}{2}\lam_T n^2$ penalizes the protocol for holding a residual
position at expiry.
It may be interpreted as the execution cost of an instantaneous block liquidation of
$n$ units at terminal impact cost $\lam_T$, or as a bequest function that assigns
negative value to unhedged residual inventory.
As $\lam_T \to \infty$, the protocol is forced to unwind fully before $T$; as
$\lam_T \to 0$, the terminal inventory is left unconstrained.
\end{remark}

\noindent
By the principle of dynamic programming, $V^T : [0,T] \times \Rb^2 \to \Rb$
satisfies the HJB PDE
\begin{align}
-\d_t V^T
    = \sup_{\gam\,\in\,\Rb}\Big\{
        \big[(q+\kappa)d - \kappa m + r + \mu\gam\big]n
        - \lam\gam^2 - \phi n^2
        + \big[-\kappa(d-m) - \mu\gam\big]\d_d V^T
        + \gam\,\d_n V^T
        + \tfrac{c^2}{2}\,\d_{dd}V^T
    \Big\},
\label{eq:HJB-FH}
\end{align}
with terminal condition \eqref{eq:terminal}.
The first-order condition for $\gam$ is identical in structure to \eqref{eq:FOC}:
\begin{align}
\mu n - 2\lam\gam - \mu\,\d_d V^T + \d_n V^T = 0
\quad\Longrightarrow\quad
\gam^*(t,d,n) = \frac{\mu n - \mu\,\d_d V^T + \d_n V^T}{2\lam},
\label{eq:gstar-FH}
\end{align}
and substituting back gives the reduced HJB:
\begin{align}
-\d_t V^T
    = \big[(q+\kappa)d - \kappa m + r\big]n - \phi n^2
      - \kappa(d-m)\,\d_d V^T
      + \frac{\big(\mu n - \mu\,\d_d V^T + \d_n V^T\big)^2}{4\lam}
      + \frac{c^2}{2}\,\d_{dd}V^T.
\label{eq:HJB-FH-reduced}
\end{align}

\subsection{Explicit Solution}
\label{sec:FH-solution}

%\subsubsection{Quadratic Ansatz and ODE System}

We seek a solution of the form
\begin{align}
V^T(t,d,n) = A(t)\,n^2 + B(t)\,nd + C(t)\,n + E(t)\,d^2 + F(t)\,d + G(t),
\label{eq:ansatz-FH}
\end{align}
with time-dependent coefficients satisfying the terminal conditions
\begin{align}
A(T) &= -\tfrac{1}{2}\lam_T, &
B(T) &= E(T) = C(T) = F(T) = G(T) = 0.
\label{eq:terminal-coeffs}
\end{align}
The partial derivatives are
\begin{align}
\d_d V^T = B n + 2E d + F, \qquad
\d_n V^T = 2A n + B d + C, \qquad
\d_{dd}V^T = 2E,
\label{eq:partials-FH}
\end{align}
and the combined numerator of \eqref{eq:gstar-FH} is
\begin{align}
\Phi^T(t,d,n)
    := \mu n - \mu\,\d_d V^T + \d_n V^T
    = \underbrace{\big(\mu(1-B) + 2A\big)}_{=:\,2\lam\Gam_N(t)}\,n
    + \underbrace{\big(B - 2\mu E\big)}_{=:\,2\lam\Gam_D(t)}\,d
    + \underbrace{\big(C - \mu F\big)}_{=:\,2\lam\Gam_0(t)},
\label{eq:Phi-FH}
\end{align}
%defining the time-dependent feedback gains
where we have defined
\begin{align}
%\Gam_N(t) &:= \frac{\mu(1-B(t)) + 2A(t)}{2\lam}, \label{eq:GamN}\\
%\Gam_D(t) &:= \frac{B(t) - 2\mu E(t)}{2\lam},    \label{eq:GamD}\\
%\Gam_0(t) &:= \frac{C(t) - \mu F(t)}{2\lam}.      \label{eq:Gam0}
\Gam_N(t) &:= \frac{\mu(1-B(t)) + 2A(t)}{2\lam}, &
\Gam_D(t) &:= \frac{B(t) - 2\mu E(t)}{2\lam},   &
\Gam_0(t) &:= \frac{C(t) - \mu F(t)}{2\lam}.     \label{eq:GamN}
\end{align}
If the value function is of the form \eqref{eq:ansatz-FH}, then
the optimal control at time $t$ is given by
\begin{align}
\gam^*(t, D_t^\gam, N_t^\gam)
    = \Gam_N(t)\,N_t^\gam + \Gam_D(t)\,D_t^\gam + \Gam_0(t).
\label{eq:gstar-FH-feedback}
\end{align}
We are now in a position to present our main result for the finite-horizon stochastic control problem.

%\subsubsection{Main Result}

%\begin{proposition}[Finite-Horizon Explicit Solution]
%\label{prop:finite}
%Under the parameter conditions of Proposition~\ref{prop:main} with $\rho = 0$,
%the finite-horizon value function \eqref{eq:V-FH} is given by the quadratic
%\eqref{eq:ansatz-FH} and the optimal control is the time-dependent linear
%feedback \eqref{eq:gstar-FH-feedback}, where the coefficient functions
%$A, B, C, E, F, G : [0,T] \to \Rb$ satisfy the following backward ODE system,
%solved on $[0,T]$ with terminal conditions \eqref{eq:terminal-coeffs}.
%\end{proposition}

\begin{proposition}[Finite-Horizon Explicit Solution]
\label{prop:finite}
Under the parameter conditions of Proposition~\ref{prop:main} with $\rho = 0$,
the finite-horizon value function \eqref{eq:V-FH} is given by the quadratic
\eqref{eq:ansatz-FH} and the optimal control is the time-dependent linear
feedback \eqref{eq:gstar-FH-feedback}, where the coefficient functions
$A, B, C, E, F, G : [0,T] \to \Rb$ satisfy the Riccati ODEs
\eqref{eq:ode-A}--\eqref{eq:ode-B}, the linear ODEs
\eqref{eq:ode-C}--\eqref{eq:ode-F} (integrated numerically as the coupled system
\eqref{eq:ode-Cc}), and the
quadrature \eqref{eq:ode-G}, all solved backward from the terminal conditions
\eqref{eq:terminal-coeffs}.
\end{proposition}

\begin{proof}
Substituting the ansatz \eqref{eq:ansatz-FH} and its partial derivatives
\eqref{eq:partials-FH} into the reduced HJB \eqref{eq:HJB-FH-reduced} and collecting
by monomial in $(d,n)$ yields the following system.

\medskip
\noindent\textit{Coefficient of $n^2$ --- Riccati ODE for $A$:}
\begin{align}
\dot{A}(t)
    = \phi - \lam\Gam_N(t)^2
    = \phi - \frac{\big[\mu(1-B) + 2A\big]^2}{4\lam},
    \qquad A(T) = -\tfrac{1}{2}\lam_T.
\label{eq:ode-A}
\end{align}

\medskip
\noindent\textit{Coefficient of $d^2$ --- Riccati ODE for $E$:}
\begin{align}
\dot{E}(t)
    = 2\kappa\,E - \lam\Gam_D(t)^2
    = 2\kappa\,E - \frac{(B - 2\mu E)^2}{4\lam},
    \qquad E(T) = 0.
\label{eq:ode-E}
\end{align}

\medskip
\noindent\textit{Coefficient of $nd$ --- Riccati ODE for $B$:}
\begin{align}
\dot{B}(t)
    = -(q+\kappa) + \kappa B
      - \frac{\big[\mu(1-B)+2A\big]\big(B - 2\mu E\big)}{2\lam},
    \qquad B(T) = 0.
\label{eq:ode-B}
\end{align}
Equations \eqref{eq:ode-A}--\eqref{eq:ode-B} form a closed Riccati system for
$(A, E, B)$, to be solved first (backward from $T$).

\medskip
\noindent\textit{Coefficient of $n$ --- linear ODE for $C$:}
\noindent
Once $(A,B,E)$ are known, collecting the $n$-coefficient gives
\begin{align}
\dot{C}(t)
    = -(r - \kappa m) - \kappa m B
      - \Gam_N(t)\big(C - \mu F\big),
    \qquad C(T) = 0.
\label{eq:ode-C}
\end{align}

\medskip
\noindent\textit{Coefficient of $d$ --- linear ODE for $F$:}
\begin{align}
\dot{F}(t)
    = \kappa F - 2\kappa m E
      - \Gam_D(t)\big(C - \mu F\big),
    \qquad F(T) = 0.
\label{eq:ode-F}
\end{align}
The quantity $\Cc(t) := C(t) - \mu F(t) = 2\lam\Gam_0(t)$.
Subtracting $\mu$ times \eqref{eq:ode-F} from \eqref{eq:ode-C} gives
\begin{align}
\dot{\Cc}(t)
    = -r + \kappa m\big(1 - B(t) + 2\mu E(t)\big)
      - \big(\Gam_N(t) - \mu\Gam_D(t)\big)\Cc(t)
      - \mu\kappa F(t),
    \qquad \Cc(T) = 0.
\label{eq:ode-Cc}
\end{align}
Because the residual $-\mu\kappa F(t)$ involves $F$ rather than $\Cc$,
the system \eqref{eq:ode-C}--\eqref{eq:ode-F} does not reduce to a scalar ODE
for $\Cc$ alone; it must be solved as the coupled linear system
\eqref{eq:ode-C}--\eqref{eq:ode-F}, which is numerically integrated alongside
the Riccati block.

\medskip
\noindent\textit{Constant term:}
\begin{align}
\dot{G}(t)
    = -\lam\Gam_0(t)^2 - c^2 E(t) - \kappa m F(t),
    \qquad G(T) = 0,
\label{eq:ode-G}
\end{align}
which is obtained by direct integration once $(E, F, \Gam_0)$ are known.
\end{proof}

\begin{remark}[Hierarchical structure of the ODE system]
\label{rem:riccati}
The backward ODE system has a hierarchical cascade mirroring the algebraic cascade in
the infinite-horizon proof.
The Riccati block $(A, E, B)$ in \eqref{eq:ode-A}--\eqref{eq:ode-B} is independent
of $(C, F, G)$ and is solved first.
The linear block $(C, F)$ is then driven by the already-solved Riccati block;
it is integrated numerically as the coupled system \eqref{eq:ode-C}--\eqref{eq:ode-F}.
Finally, $G$ follows by quadrature from \eqref{eq:ode-G}.
\end{remark}

\begin{remark}[Special case: no permanent impact: $\mu = 0$]
\label{rem:mu0}
When $\mu = 0$ the system becomes triangular rather than fully decoupled.
The ODE for $A$ reduces to the scalar Riccati equation $\dot{A} = \phi - A^2/\lam$
with $A(T) = -\frac{1}{2}\lam_T$.
Setting $a := \sqrt{\lam\phi}$ and $s := \sqrt{\phi/\lam}\,(T-t)$, the solution satisfying the terminal condition is
\begin{align}
A(t) = -a\,
    \frac{\lam_T/2 + a\tanh s}
         {a + (\lam_T/2)\tanh s}.
\label{eq:A-mu0}
\end{align}
One can verify directly that $A(T) = -\lam_T/2$.
Given $A(t)$, the ODE for $B$ when $\mu=0$ reads
\begin{align}
\dot{B} = -(q+\kappa) + \kappa B - \frac{A(t)}{\lam}\,B, \qquad B(T) = 0,
\label{eq:B-mu0-ode}
\end{align}
which is a linear (non-autonomous) first-order ODE and must be integrated numerically in general.
The two ODEs for $E$ and the linear coefficients $(C, F)$ may then be solved in sequence.
\end{remark}

\subsection{Economic Implications}
\label{sec:FH-discussion}

The finite-horizon solution shares several qualitative features with the
infinite-horizon solution of Section~\ref{sec:IH}, but the time-dependence of the
feedback coefficients $\Gam_N(t)$, $\Gam_D(t)$, $\Gam_0(t)$ introduces important new phenomena.

\paragraph{Liquidation urgency near expiry.}
As $t \to T$, the terminal condition $A(T) = -\frac{1}{2}\lam_T < 0$ pulls
$\Gam_N(t)$ sharply negative.
From \eqref{eq:GamN} with $B(T)=0$ and $A(T)=-\lam_T/2$, we have
$\Gam_N(T) = (\mu - \lam_T)/(2\lam)$.
This is negative whenever $\lam_T > \mu$, an inequality we assume henceforth
(it holds for the numerical parameters of Section~\ref{sec:numerical}).
Under this condition, near $T$ the control satisfies
$\gam^* \approx \big[(\mu-\lam_T)/(2\lam)\big] N_t^\gam$ --- a rate of unwinding
proportional to the current position size.
Note that the terminal penalty $\frac{1}{2}\lam_T(N_T^\gam)^2$ incentivizes
liquidation but does not enforce $N_T^\gam = 0$; for a hard terminal constraint
one would take $\lam_T \to \infty$.
This liquidation urgency is entirely absent from the infinite-horizon policy and
reflects the finite deadline.

\paragraph{Basis chasing early, deleveraging late.}
Far from expiry ($T - t$ large), the time-dependent feedback coefficients stabilize
toward their long-run (undiscounted stationary) values.
Because the finite-horizon problem is undiscounted ($\rho = 0$), these long-run values
differ in general from the discounted infinite-horizon coefficients $\gam_D$, $\gam_N$
of Proposition~\ref{prop:main}; see the ``Relation to the infinite-horizon solution''
paragraph below.
Nevertheless, the qualitative policy---basis-chasing combined with inventory-risk
braking---is the same when time to expiry is long.
As $t \to T$, the basis-chasing feedback coefficient $\Gam_D(t)$ diminishes because the remaining
horizon over which future funding can be collected shrinks, making basis-driven
accumulation less attractive relative to the cost of the terminal liquidation penalty.

\paragraph{Interaction between terminal cost and inventory penalty.}
The terminal cost $\lam_T$ and the running inventory penalty $\phi$ both discourage
large positions, but through different channels.
The running penalty $\phi$ operates continuously, slowing accumulation throughout
$[0, T]$.
The terminal cost $\lam_T$ operates only at $T$ but propagates backward through the
Riccati ODE \eqref{eq:ode-A}, shaping the entire time profile of $\Gam_N(t)$.
A large $\lam_T$ (high liquidation cost) induces early de-risking, effectively
front-loading the liquidation and spreading it over $[0,T]$ rather than concentrating
it at $T$.

\paragraph{Relation to the infinite-horizon solution.}
The finite-horizon problem is formulated without discounting ($\rho = 0$),
whereas the infinite-horizon problem of Section~\ref{sec:IH} uses $\rho > 0$.
Accordingly, the two solutions do not directly nest one inside the other:
the stationary limits of $A(t)$, $B(t)$, $E(t)$ as $T\to\infty$ satisfy the
algebraic equations of the undiscounted problem, not those of
Proposition~\ref{prop:main}.
Nevertheless, the two solutions share the same qualitative structure, and for
long-lived protocols the infinite-horizon feedback coefficients
$(\gam_N, \gam_D, \gam_0)$ provide a tractable and economically interpretable
approximation when the remaining horizon is large.

\paragraph{Closed-loop dynamics.}
Under the optimal control \eqref{eq:gstar-FH-feedback}, the state processes satisfy
the two-dimensional linear SDE
\begin{align}
\dd D_t^\gam
    &= \big[-(\kappa + \mu\Gam_D(t))\,D_t^\gam
            - \mu\Gam_N(t)\,N_t^\gam
            + \kappa m - \mu\Gam_0(t)\big]\dd t + c\,\dd W_t,
\label{eq:dD-closed-FH}\\
\dd N_t^\gam
    &= \big[\Gam_N(t)\,N_t^\gam + \Gam_D(t)\,D_t^\gam + \Gam_0(t)\big]\dd t,
\label{eq:dN-closed-FH}
\end{align}
whose solution is Gaussian for all $t \in [0,T]$.
The time-varying drift matrix
\begin{align}
M(t) :=
\begin{pmatrix}
-(\kappa + \mu\Gam_D(t)) & -\mu\Gam_N(t) \\
\Gam_D(t) & \Gam_N(t)
\end{pmatrix}
\label{eq:drift-matrix-FH}
\end{align}
has the same qualitative structure as the infinite-horizon matrix \eqref{eq:drift-matrix},
with time-independent feedback coefficients replaced by time-dependent ones that stabilize
toward their long-run (undiscounted stationary) values as $T - t \to \infty$.

%-----------------------------------------------------------------------------------
%
%       Numerical Experiments
%
%-----------------------------------------------------------------------------------

\section{Numerical Experiments}
\label{sec:numerical}
\begin{figure}
\begin{tabular}{ccc}
\includegraphics[width=0.310\textwidth]{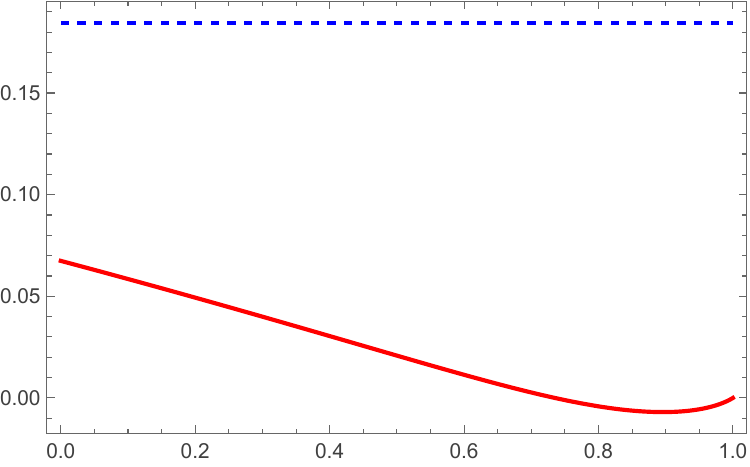}&
\includegraphics[width=0.310\textwidth]{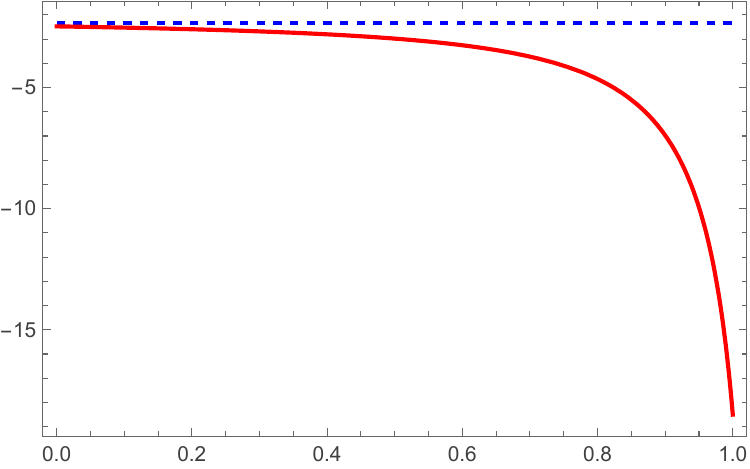}&
\includegraphics[width=0.310\textwidth]{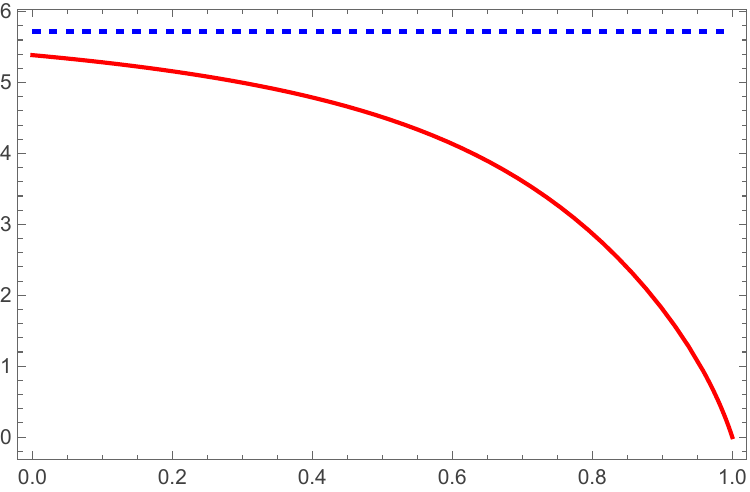}\\
$\Gam_0(t)$ and $\gam_0$ & $\Gam_N(t)$ and $\gam_N$ &  $\Gam_D(t)$ and $\gam_D$
\end{tabular}
\caption{Using the model parameters given in Section \ref{sec:numerical} we plot $\Gam_0(t)$, $\Gam_N(t)$ and $\Gam_D(t)$, given in \eqref{eq:GamN}, as functions of $ t$ (red lines).  For comparison, we also plot $\gam_0$, $\gam_N$ and $\gam_D$, given in \eqref{eq:gamN} (dashed blue lines).}
\label{fig:gamma}
\end{figure}

In this section, we illustrate the optimal Ethena strategy by comparing the time-dependent finite-horizon policy with the stationary infinite-horizon policy. To highlight the qualitative differences between these two regimes, we use the following model parameters:
\begin{align}
\rho &= 0.05, &
\kappa &= 2.0, &
m &= 0.04, &
\mu &= 0.3, &
\lam &= 0.1,  &
r &= 0.04,  & 
q &= 4.0,    \\  
c& = 0.1,     &  
\phi &= 0.5,   &  
\lam_T &= 4.0, &
T &= 1.0, &
D_0^\gam	&=	m , &
N_0^\gam &= 0.0 , &
X_0^\gam &= 0.0 .
\end{align}
In Figure \ref{fig:gamma}, we plot $\Gam_0(t)$, $\Gam_N(t)$ and $\Gam_D(t)$ as a function of $t$ (solid red lines).  For comparison, we also plot the constants $\gam_0$, $\gam_N$ and $\gam_D$ (dashed blue lines).  Several key behaviors emerge:
\begin{itemize}
\item 
\textit{Qualitative similarity far from expiry:} For times $t$ far from the terminal date $T$, the finite-horizon coefficients $(\Gam_0(t),\Gam_N(t),\Gam_D(t))$ are close to the discounted infinite-horizon values $(\gam_0,\gam_N,\gam_D)$ for the parameter values used here. As noted in the ``Relation to the infinite-horizon solution'' paragraph of Section~\ref{sec:FH}, the two problems use different discount rates ($\rho = 0$ versus $\rho > 0$), so this near-coincidence is a feature of the chosen parameters rather than a general identity. Nevertheless, for a protocol with a long operational runway, Figure~\ref{fig:gamma} suggests that the stationary infinite-horizon policy provides a qualitatively accurate approximation of the optimal strategy.
\item
\textit{Liquidation Urgency:} As $t \to T$, we observe a sharp divergence in the coefficients. Most notably, $\Gam_N(t)$ becomes significantly more negative. This reflects a ``liquidation urgency''; the high terminal cost $\lambda_T$ incentivizes the protocol to begin unwinding its position well before the deadline to avoid a prohibitive lump-sum execution cost at time $T$.
\item
\textit{Basis Sensitivity:} The coefficient $\Gam_D(t)$, which governs the protocol's response to the basis (funding-rate chasing), remains relatively stable until the final stages of the horizon. This indicates that the protocol continues to harvest carry from basis fluctuations even late in its lifecycle, though this response is eventually dampened by the overarching requirement to reach a zero position.
\end{itemize}

\begin{figure}
\begin{tabular}{ccc}
\includegraphics[width=0.310\textwidth]{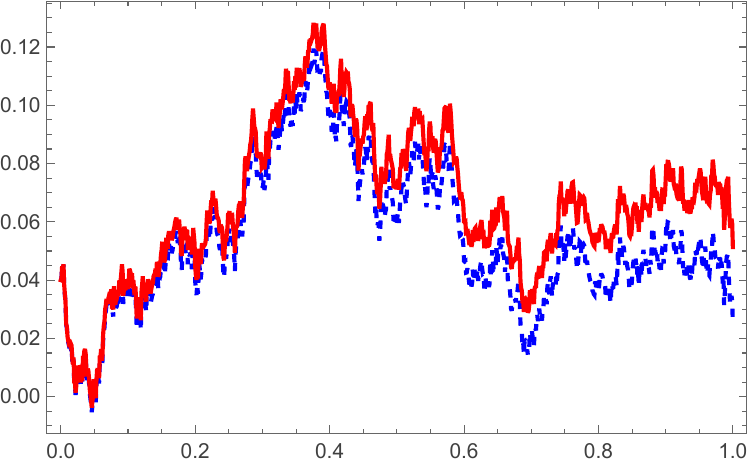}&
\includegraphics[width=0.310\textwidth]{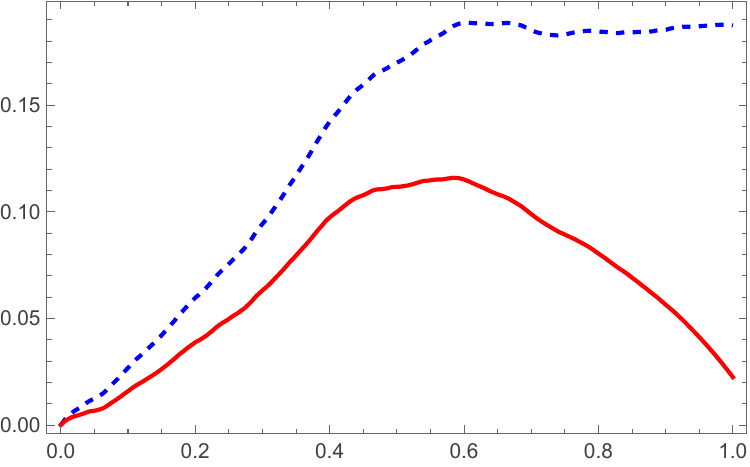}&
\includegraphics[width=0.310\textwidth]{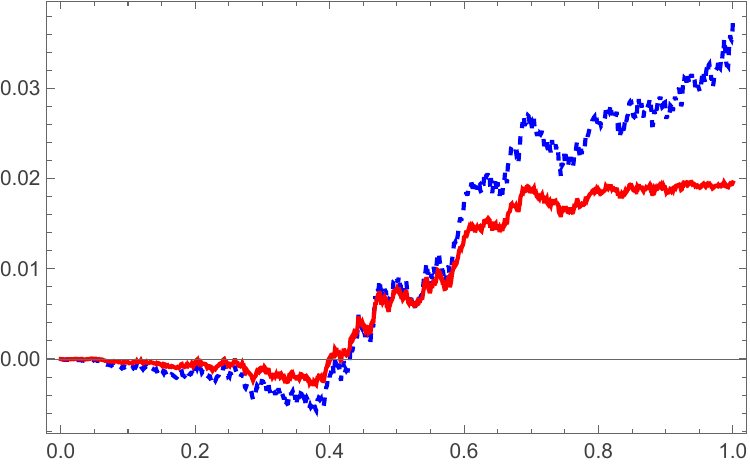}\\
$D_t^\gam$ & $N_t^\gam$ & $X_t^\gam$
\end{tabular}
\caption{Using the model parameters given in Section \ref{sec:numerical} we plot one sample path of $D_t^\gam$, $N_t^\gam$ and $X_t^\gam$ for both the finite horizon case (red lines) and the infinite horizon case (dashed blue lines).  Recall that the closed-loop dynamics of $D_t^\gam$ and $N_t^\gam$ in the infinite horizon case are given by \eqref{eq:dD-closed} and \eqref{eq:dN-closed}, respectively, while the closed-loop dynamics of $D_t^\gam$ and $N_t^\gam$ in the finite horizon case are given by \eqref{eq:dD-closed-FH} and \eqref{eq:dN-closed-FH}, respectively.  In both the infinite and finite horizon cases the dynamics of $X_t^\gam$ are given by \eqref{eq:dX} with $\gam_t$ given by \eqref{eq:gstar} and \eqref{eq:gstar-FH-feedback} in the infinite horizon and finite horizon cases, respectively.}
\label{fig:DNX}
\end{figure}

\noindent
In Figure \ref{fig:DNX}, we plot one sample path of $D_t^\gam$, $N_t^\gam$ and $X_t^\gam$ for both the finite horizon case (solid red lines) and infinite horizon case (dashed blue lines).  We note the following:
\begin{itemize}
\item
\textit{Position Smoothing:} Under the finite-horizon policy, the position $N_t^\gam$ (red) begins to deviate from the infinite-horizon path (blue) as $t$ approaches $T$. While the infinite-horizon strategy—lacking a deadline—continues to maintain a large short position to harvest carry, the finite-horizon strategy yields a smoother transition toward zero.
\item
\textit{Wealth Trade-offs:} The cumulative wealth $X_t^\gam$ highlights the internalized cost of terminal liquidation. The red path demonstrates how the protocol ``pays'' for its liquidation early through reduced accumulation rates. This proactive unwinding ensures the protocol does not face a catastrophic wealth drop at time $T$ due to the terminal penalty $\frac{1}{2}\lambda_T (N_T^\gam)^2$.
\item
\textit{Basis Interaction:} The basis paths $D_t^\gam$ remain closely coupled for much of the duration. However, as $t \to T$, the finite-horizon path experiences less self-induced compression. By reducing its trading rate near the horizon, the protocol allows the basis to mean-revert more naturally toward $m$ without the persistent downward pressure exerted by the protocol's own short-selling.
\end{itemize}

\noindent
Mathematica code that was used to create the plots is provided in Appendix \ref{sec:mathematica}.

%-----------------------------------------------------------------------------------
%
%       Stress Test
%
%-----------------------------------------------------------------------------------

\subsection{Stress Test}
\label{sec:stress}

\begin{figure}
\begin{tabular}{ccc}
\includegraphics[width=0.310\textwidth]{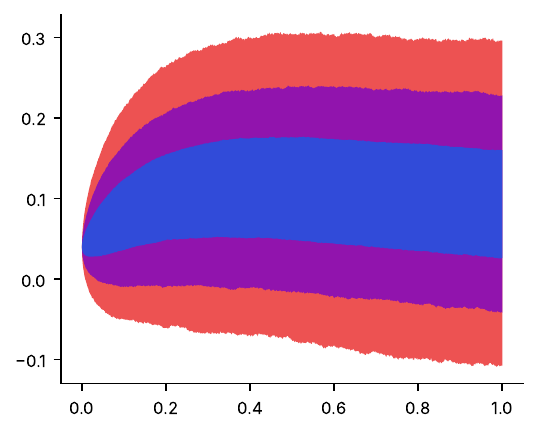}&
\includegraphics[width=0.310\textwidth]{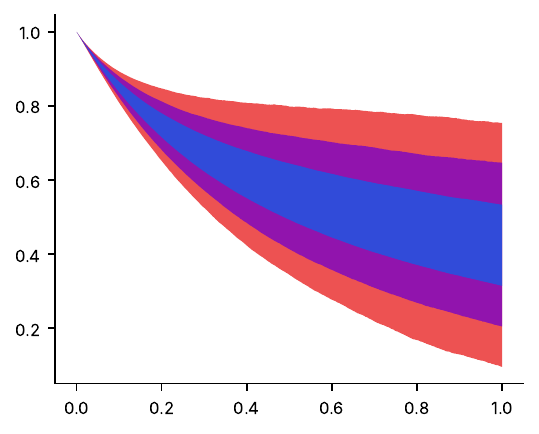}&
\includegraphics[width=0.310\textwidth]{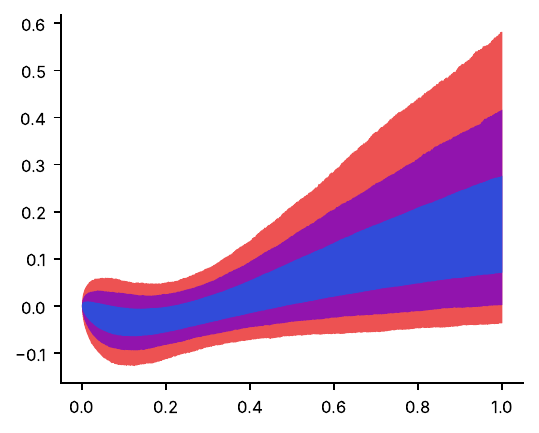}\\
$D_t^\gam$ & $N_t^\gam$ & $X_t^\gam$
\end{tabular}
\caption{Stress test using the model parameters of Section~\ref{sec:numerical}.
The infinite-horizon feedback coefficients $(\gam_N,\gam_D,\gam_0)$ are computed
once under the baseline (calm) volatility $c=0.1$ and then applied without
re-optimization to $10{,}000$ simulated paths at each volatility level.
Each shaded band shows the 5th--95th percentile of the path distribution.
Blue band: calm regime ($c$); purple band: first stress level ($2c$); red band:
second stress level ($3c$).}
\label{fig:stress}
\end{figure}

We now assess the out-of-sample robustness of the infinite-horizon policy calibrated to the
calm-regime parameters of Section~\ref{sec:numerical}.  The experiment is as follows.
We compute the infinite-horizon feedback coefficients $(\gam_N, \gam_D, \gam_0)$ once, using
the baseline volatility $c = 0.1$.  We then apply this fixed policy in two stressed environments
in which the basis volatility is scaled to $2c = 0.2$ and $3c = 0.3$, while holding all other
parameters unchanged.  Because the policy coefficients are not re-optimized for the stressed
environment, the simulation directly measures robustness to a volatility shock.
\\[0.5em]
The resulting confidence bands for $D_t^\gam$, $N_t^\gam$, and $X_t^\gam$ are displayed in
Figure~\ref{fig:stress}.  Each panel shows the 5th--95th percentile bands across 10{,}000
simulated paths: calm regime (blue, $c$), first stress level (purple, $2c$), and second
stress level (red, $3c$).  Because the policy coefficients are held fixed at their
calm-regime values, the simulation directly measures the distribution of outcomes under
a volatility shock.
\\[0.5em]
Several features emerge from the stress test:
\begin{itemize}
\item
\textit{Basis volatility:}  The 5th--95th percentile band for $D_t^\gam$ widens
proportionally with the volatility scaling, as expected from the SDE \eqref{eq:dD}.
Even at $3c$ the entire band remains mean-reverting and does not diverge, confirming
that the stability condition $\kappa + \mu\gam_D > 0$ (which depends only on the policy
and the mean-reversion speed, not on $c$) is unaffected by the volatility shock.

\item
\textit{Position dynamics:}  Because the position $N_t^\gam$ evolves through the ODE
$\dot{N}_t = \gam_N N_t + \gam_D D_t + \gam_0$ (equation~\eqref{eq:dN-closed}), its
variability enters only through the basis path $D_t^\gam$.  The confidence band for $N_t^\gam$
widens modestly under stressed volatility---the policy accumulates faster on paths where
the basis is elevated and more slowly where the basis dips---but the qualitative
shape, a smooth ramp toward a stationary level, is preserved across all three regimes.

\item
\textit{Wealth path:}  The confidence band for $X_t^\gam$ broadens significantly in the
stressed regime because the stochastic integral term $-c\,N_t^\gam\,\dd W_t$ in
\eqref{eq:dX} scales with $c$.  The simulated mean and median of $X_T^\gam$ are positive
in all three regimes (Table~\ref{tab:stress}), confirming that the calm-calibrated policy
harvests carry on average even under doubled or tripled volatility.  However, the downside
tail grows with $c$: the fraction of paths with $X_T^\gam < 0$ rises from $0.05\%$ in the
calm regime to $4.7\%$ at $2c$ and $11.0\%$ at $3c$.
\end{itemize}

\begin{table}[h]
\centering
\begin{tabular}{lccc}
\hline
Regime & Mean $X_T^\gam$ & Median $X_T^\gam$ & $\Pb(X_T^\gam < 0)$ \\
\hline
$c$   & 0.167 & 0.164 & $0.05\%$ \\
$2c$  & 0.181 & 0.165 & $4.7\%$  \\
$3c$  & 0.212 & 0.176 & $11.0\%$ \\
\hline
\end{tabular}
\caption{Simulated terminal wealth statistics across 10{,}000 paths for each volatility
regime. The mean and median are positive in all three cases; the probability of a loss
grows with the volatility scaling factor.}
\label{tab:stress}
\end{table}

\noindent
Overall, the stress test shows that the simulated wealth distributions exhibit increasing
downside risk as volatility rises, while the mean and median remain positive across all
three regimes.  The position and wealth distributions preserve their characteristic shapes,
and no instability or runaway behavior is observed.

%-----------------------------------------------------------------------------------
%
%       Conclusion
%
%-----------------------------------------------------------------------------------

\section{Conclusion and Future Research}
\label{sec:conclusion}
In this paper, we have developed a rigorous stochastic control framework for the optimal management of the Ethena protocol's delta-neutral carry strategy. By modeling the simultaneous execution of staked asset purchases and perpetual futures shorting, we derived optimal feedback policies for both infinite and finite horizons. Our results demonstrate that the optimal strategy is characterized by a balance between three competing forces: the desire to harvest staking rewards and funding income, the need to minimize permanent and temporary market impact, and the mitigation of inventory risk.
%\\[0.5em]
%A key numerical finding is the convergence of the finite-horizon policy to the stationary infinite-horizon policy when the time to maturity is large. However, as the horizon approaches, the protocol exhibits a marked ``liquidation urgency,'' proactively unwinding positions to avoid prohibitive terminal execution costs. This suggests that for an ongoing protocol, the stationary gains provide a robust and tractable heuristic, while the finite-horizon model is essential for discrete-time fund lifecycles or scheduled deleveraging events.
\\[0.5em]
There are several promising directions for future research:
\begin{itemize}
    \item \textit{Endogenous Inflows and Outflows:} Perhaps the most significant extension involves modeling the Ethena position $N_t$ not as a pure control variable, but as a process driven by user deposits and withdrawals. In the current model, the controller chooses the rate of change $\gamma_t$ to maximize wealth. In reality, the protocol must execute trades in response to minting and redeeming demand for the stablecoin. Formulating this as a tracking problem—where the protocol seeks to maintain a delta-neutral hedge while minimizing the impact costs associated with stochastic AUM fluctuations—would be of great practical interest.
    \item \textit{Crowding effect and endogenous long-run basis:} In the current model, the long-run basis level $m$ is exogenous. In practice, the impact of order flow $\gamma_t$ on the basis enters through the drift of an OU process and decays over time at rate $\kappa$, so the basis returns to $m$ regardless of what the protocol has done. The persistent effect Ethena actually has on funding is a crowding effect: because the protocol constitutes a large share of cash-and-carry arbitrage capital, the equilibrium funding level depends on the protocol's cumulative position $N$, not the trading rate $\gamma$. Replacing $m$ by $m - \beta N$ in \eqref{eq:dD} (with $\beta > 0$) would capture this crowding effect while preserving the linear-quadratic structure of the problem and yielding an analytically tractable model of essentially the same form.
    \item \textit{Two-factor basis dynamics:} A single Ornstein--Uhlenbeck factor for the basis conflates two empirically distinct timescales: the premium index, which reverts within hours, and the broader funding level, which evolves over weeks. A two-factor specification $D_t = X_t^1 + X_t^2$, with two independent OU components at different mean-reversion speeds, would better capture this structure. The state becomes $(X_t^1, X_t^2, N_t) \in \Rb^3$, which remains in the linear-quadratic framework. The value function is then a quadratic in three variables with ten coefficients, and the HJB equation again reduces to a cascade of Riccati equations --- a larger but still explicit algebraic system in the infinite-horizon case, and a larger but still tractable system of Riccati ODEs in the finite-horizon case.
    \item \textit{Stochastic Parameters and Regime Switching:} While we modeled the basis as a mean-reverting process, the staking rewards $r$ and the funding sensitivity $q$ were treated as constants. Future work could incorporate stochastic volatility or regime-switching dynamics to capture ``funding rate spikes'' or periods of extreme market stress where the basis deviates significantly from historical norms.
    \item \textit{Multi-Asset Portfolio Optimization:} Ethena has expanded to include assets beyond Ethereum, such as Bitcoin and Solana. Extending this framework to a multi-asset setting would allow for the study of optimal collateral diversification and the cross-impact effects of simultaneous hedging across multiple perpetual markets.
    \item \textit{Counterparty and Protocol Risk:} Integrating a ``jump-to-default'' or a credit risk component to account for the potential failure of centralized exchanges or the de-pegging of the staked asset would provide a more comprehensive view of the protocol's risk-adjusted returns.
\end{itemize}

\section*{Acknowledgments}
We would like to thank an anonymous referee for a careful reading of this manuscript and for comments that improved the paper.

%-----------------------------------------------------------------------------------
%
%       Mathematica Code
%
%-----------------------------------------------------------------------------------

\appendix
\section{\textit{Wolfram Mathematica} \textcopyright{} code}
\label{sec:mathematica}
In this Appendix, we provide the \textit{Wolfram Mathematica} \textcopyright{} code that was used to produce the plots in Section \ref{sec:numerical}.
\\{$\,$}
\begin{lstlisting}
(* 1. MODEL PARAMETERS *)
(* These parameters are chosen to satisfy the stability condition for the infinite horizon *)
rho = 0.05;    (* Discount rate *)
kappa = 2.0;   (* Basis mean-reversion speed (x in the model) *)
m = 0.04;      (* Long-run basis level *)
mu = 0.3;      (* Total permanent impact *)
lambda = 0.1;  (* Total temporary impact *)
r = 0.04;      (* Staking yield *)
q = 4.0;       (* Funding rate factor *)
c = 0.1;       (* Basis volatility *)
phi = 0.5;     (* Inventory risk penalty *)
lambdaT = 4.0; (* Terminal liquidation cost *)
Tmax = 1.0;    (* Time horizon *)

xStar = rho + 2*kappa;

(* 2. INFINITE HORIZON SOLUTION *)
a1Inf[a2_] := (lambda*rho - mu*(1 - a2) - Sqrt[(lambda*rho - mu*(1 - a2))^2 + 4*lambda*phi - mu^2*(1 - a2)^2])/2;
a4Inf[a2_] := (mu*a2 + lambda*xStar - Sqrt[lambda*xStar*(lambda*xStar + 2*mu*a2)])/(2*mu^2);

valA2 = a2 /. FindRoot[(rho + kappa)*a2 == (q + kappa) + ((mu*(1 - a2) + 2*a1Inf[a2])*(a2 - 2*mu*a4Inf[a2]))/(2*lambda), {a2, 0}];

infA1 = a1Inf[valA2];
infA4 = a4Inf[valA2];
infA2 = valA2;

gammaNInf = (mu*(1 - infA2) + 2*infA1)/(2*lambda);
gammaDInf = (infA2 - 2*mu*infA4)/(2*lambda);
(* 2x2 Cramer's rule for alpha3, alpha5, and C = alpha3 - mu*alpha5 *)
rInf = (r - kappa*m) + kappa*m*infA2;
sInf = 2*kappa*m*infA4;
deltaInf = (rho - gammaNInf)*(rho + kappa) + rho*mu*gammaDInf;
cInf = (rInf*(rho + kappa) - rho*mu*sInf)/deltaInf;
alpha3Inf = (rInf + gammaNInf*cInf)/rho;
alpha5Inf = (alpha3Inf - cInf)/mu;
gamma0Inf = cInf/(2*lambda);

(* 3. FINITE HORIZON SOLUTION *)
finiteODE = NDSolve[{
    a'[t] == phi - (mu*(1 - b[t]) + 2*a[t])^2/(4*lambda),
    e'[t] == 2*kappa*e[t] - (b[t] - 2*mu*e[t])^2/(4*lambda),
    b'[t] == -(q + kappa) + kappa*b[t] - (mu*(1 - b[t]) + 2*a[t])*(b[t] - 2*mu*e[t])/(2*lambda),
    (* Coupled linear system for cv[t]=C(t) and ff[t]=F(t) *)
    cv'[t] == -(r - kappa*m) - kappa*m*b[t] - (mu*(1 - b[t]) + 2*a[t])/(2*lambda)*(cv[t] - mu*ff[t]),
    ff'[t] == kappa*ff[t] - 2*kappa*m*e[t] - (b[t] - 2*mu*e[t])/(2*lambda)*(cv[t] - mu*ff[t]),
    a[Tmax] == -lambdaT/2, b[Tmax] == 0, e[Tmax] == 0, cv[Tmax] == 0, ff[Tmax] == 0
}, {a, b, e, cv, ff}, {t, 0, Tmax}][[1]];

GammaN[t_] := (mu*(1 - b[t]) + 2*a[t])/(2*lambda) /. finiteODE;
GammaD[t_] := (b[t] - 2*mu*e[t])/(2*lambda) /. finiteODE;
Gamma0[t_] := (cv[t] - mu*ff[t])/(2*lambda) /. finiteODE;

(* 4. PLOTS OF FEEDBACK CONTROL COEFFICIENTS *)
Print["Comparison of constant feedback coefficient gamma_0 (Blue, Dashed) and time-dependent feedback coefficient Gamma_0(t) (Red, Solid)"];
Plot[{gamma0Inf, Gamma0[t]}, {t, 0, Tmax}, PlotStyle -> {{Blue, Dashed}, {Red}}, Frame -> True, PlotRange -> Full]

Print["Comparison of constant feedback coefficient gamma_N (Blue, Dashed) and time-dependent feedback coefficient Gamma_N(t) (Red, Solid)"];
Plot[{gammaNInf, GammaN[t]}, {t, 0, Tmax}, PlotStyle -> {{Blue, Dashed}, {Red}}, Frame -> True, PlotRange -> Full]

Print["Comparison of constant feedback coefficient gamma_D (Blue, Dashed) and time-dependent feedback coefficient Gamma_D(t) (Red, Solid)"];
Plot[{gammaDInf, GammaD[t]}, {t, 0, Tmax}, PlotStyle -> {{Blue, Dashed}, {Red}}, Frame -> True, PlotRange -> Full]

(* 5. SIMULATION *)
dt = 0.001; steps = Round[Tmax/dt];
times = Table[i*dt, {i, 0, steps}];
dW = RandomVariate[NormalDistribution[0, Sqrt[dt]], steps];

(* Paths: [[1]] is Infinite Horizon (Blue), [[2]] is Finite Horizon (Red) *)
pathD = {{m}, {m}}; pathN = {{0.}, {0.}}; pathX = {{0.}, {0.}};

Do[
    currT = (i - 1)*dt;
    
    (* Controls *)
    gInf = gammaNInf*Last[pathN[[1]]] + gammaDInf*Last[pathD[[1]]] + gamma0Inf;
    gFin = GammaN[currT]*Last[pathN[[2]]] + GammaD[currT]*Last[pathD[[2]]] + Gamma0[currT];
    
    (* Dynamics for Infinite Horizon *)
    AppendTo[pathD[[1]], Last[pathD[[1]]] + (-kappa*(Last[pathD[[1]]] - m) - mu*gInf)*dt + c*dW[[i]]];
    AppendTo[pathN[[1]], Last[pathN[[1]]] + gInf*dt];
    AppendTo[pathX[[1]], Last[pathX[[1]]] + ((q + kappa)*pathD[[1]][[-2]] - kappa*m + r + mu*gInf)*pathN[[1]][[-2]]*dt - lambda*gInf^2*dt - c*pathN[[1]][[-2]]*dW[[i]]];
    
    (* Dynamics for Finite Horizon *)
    AppendTo[pathD[[2]], Last[pathD[[2]]] + (-kappa*(Last[pathD[[2]]] - m) - mu*gFin)*dt + c*dW[[i]]];
    AppendTo[pathN[[2]], Last[pathN[[2]]] + gFin*dt];
    AppendTo[pathX[[2]], Last[pathX[[2]]] + ((q + kappa)*pathD[[2]][[-2]] - kappa*m + r + mu*gFin)*pathN[[2]][[-2]]*dt - lambda*gFin^2*dt - c*pathN[[2]][[-2]]*dW[[i]]];
, {i, 1, steps}];

Print["(i) Basis paths D_t: Infinite (Blue, Dashed) vs Finite (Red, Solid)"];
ListLinePlot[{Transpose[{times, pathD[[1]]}], Transpose[{times, pathD[[2]]}]}, PlotStyle -> {{Blue, Dashed}, {Red}}, Frame -> True, PlotRange -> Full]

Print["(ii) Position paths N_t: Infinite (Blue, Dashed) vs Finite (Red, Solid)"];
ListLinePlot[{Transpose[{times, pathN[[1]]}], Transpose[{times, pathN[[2]]}]}, PlotStyle -> {{Blue, Dashed}, {Red}}, Frame -> True, PlotRange -> Full]

Print["(iii) Net Wealth paths X_t: Infinite (Blue, Dashed) vs Finite (Red, Solid)"];
ListLinePlot[{Transpose[{times, pathX[[1]]}], Transpose[{times, pathX[[2]]}]}, PlotStyle -> {{Blue, Dashed}, {Red}}, Frame -> True, PlotRange -> Full]

\end{lstlisting}
% Ethana-1.00-test.nb

%-----------------------------------------------------------------------------------
%
%       BIBLIOGRAPHY
%
%-----------------------------------------------------------------------------------

\clearpage

\bibliography{references}

%-----------------------------------------------------------------------------------
%
%       FIGURES
%
%-----------------------------------------------------------------------------------

\clearpage

% Ethana-2.00-test.nb

% Ethana-2.00-test.nb

% stress-test.nb

\end{document}